\documentclass[12pt]{iopart}
\usepackage{graphicx}

\begin{document}

\title[Dimensional Analysis and Rutherford Scattering]{Dimensional Analysis and Rutherford Scattering}

\author{M A Bernal$ˆ1$, F J Camacho$ˆ2$,
R E Martinez$ˆ2$}

\address{$ˆ1$Departamento de Ciencias B\'asicas, Polit\'ecnico Grancolombiano, Bogota, Colombia.}

\address{$ˆ2$ Departamento de F\'isica, Universidad Nacional de Colombia, Bogota, Colombia}

\ead{fjcamachor@unal.edu.co}

\begin{abstract}
Dimensional analysis, and in particular the Buckingham $\Pi$ theorem is widely used in fluid mechanics. In this article we obtain an expression for the impact parameter from Buckingham's theorem and we compare our result with  Rutherford's original discovery found in the early twentieth century.
\end{abstract}


\submitto{\EJP}

\maketitle

\section{Rutherford Scattering}
\noindent In his work about Helium nuclei scattered by a heavy nucleus, Rutherford had to consider how much close this beam of nuclei approach to the heavy target. Obviously he was not able to see what happens at microscopic scales when the scattering of one single helium atom by the effect of a nucleus of the target occurs, so he got interested into macroscopic measurable quantities. One of this quantities is the differential cross section $\sigma$ defined as the number of scattered particles per unit of time in a solid angle and divided by the incident intensity of the alpha particles beam \cite{1} \cite{2}.

In order to be able to compute $\sigma$, he needed to find a relation between the impact parameter $s$ of a single alpha particle and the scattering angle $\Phi$ (See Figure 1). $\sigma$ depends on the impact parameter and the  scattering angle as follows:

\begin{equation}
\sigma=\frac{s}{\sin\Phi}\arrowvert \frac{ds}{d\Phi} \arrowvert.
\end{equation}

Using the orbit equation of a particle under the effect of a central force, like the Coulomb interaction, is possible to obtain $s=s(\Phi)$. Following this procedure the desired relation, after some algebra,  turns out to be \cite{2}: 

\begin{equation}
s=\frac{kZZ'e^2}{2E}\cot\frac{\Phi}{2}.
\end{equation}

Here $Z$ and $Z'$ are the atomic numbers of the heavy nucleus and the incident alpha particle respectively, $k$ is the force constant of electrostatic Coulomb interaction, $E=m_{\alpha}v^2/2$ is the energy of the alpha particle and $e$ is the electron charge.

\begin{figure}[htp]
	\centering 
		\includegraphics[scale=0.45]{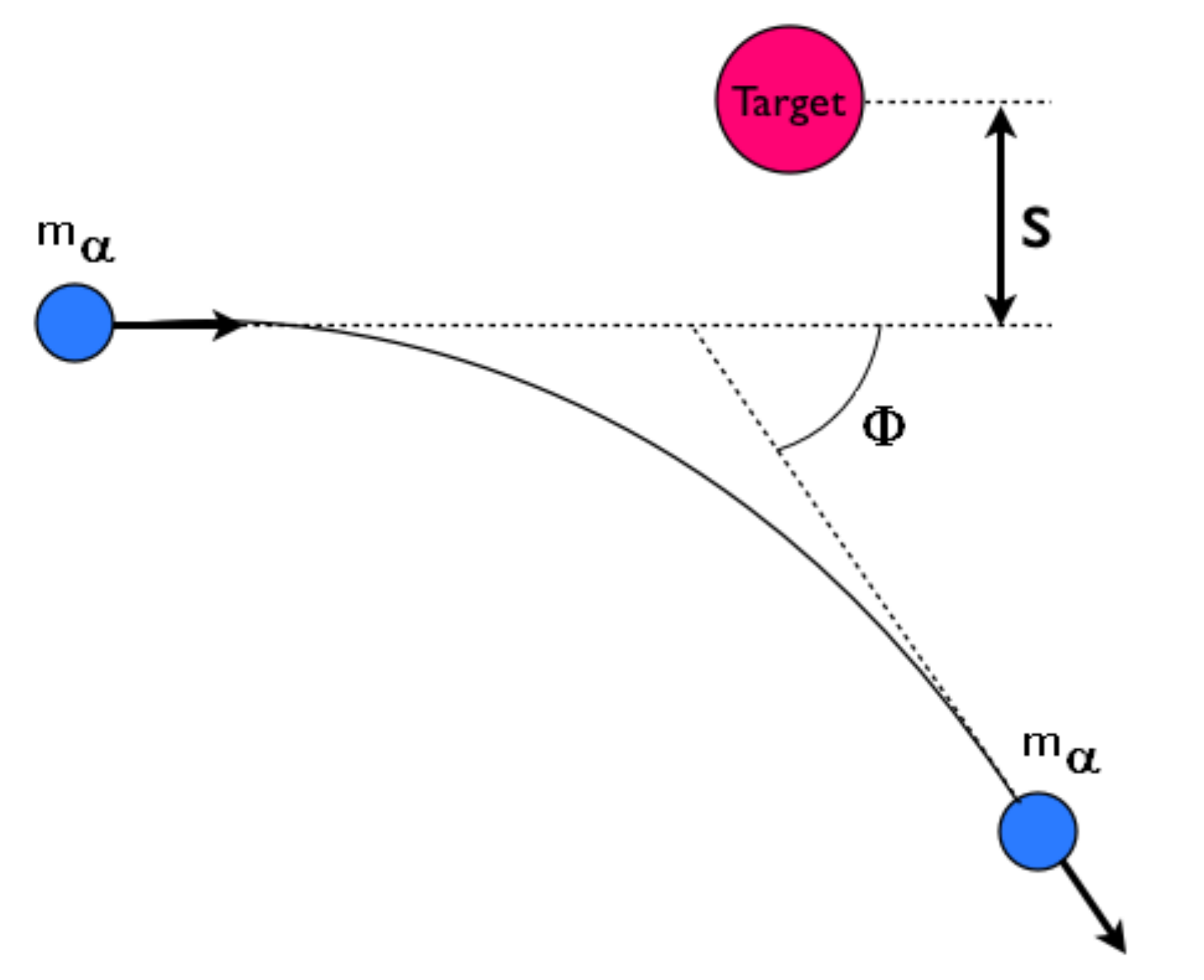}
\caption{Scheme of the Rutherford scattering. An alpha particle is fired close to the target, then due to the electrostatic interaction it gets scattered and change its trayectory by an angle $\Phi$.}
	\label{fig:un}
\end{figure}

\section{Dimensional Analysis}

In 1968 the scientific comunity established the International System of Units (IS). Constituted by the so called base units, such that any other posible unit that might describe any physical process could be derived from these base units. 
In physics an unit system like the IS is required for the dimensional analysis, which plays a fundamental role in many branches of physics. Dimensional analysis is the one in charge to verify that the equations describing physical phenomena must be dimensionally homogeneous.

As a simple example consider the Bernoulli equation, which describes the energy conservation law for a inviscid fluid:

\begin{equation}
P_1+\frac{1}{2}\rho v_1^2+\rho g h_1= P_2+\frac{1}{2}\rho v_2^2+\rho g h_2.
\end{equation} 

\noindent The subindex $1$ o $2$ indicates that the quantity is taken in two diferent points inside the fluid. $P$ is the local pressure, $v$ the velocity, $\rho$ is the fluid density and $h$ is the local height of the fluid.

Is easy to see using dimensional analysis that the dimensions of the above equation are given by $ML^{-1}T^{-2}$, where $M$ is a dimension of mass, $L$ the dimension of length and $T$ the dimension of time. This holds for both sides of (3) satisfying the requirement of dimensional homogeneity; the interesting feature of this kind of mathematical expressions is that they can be rewrittren in a dimensionless form by multiplying by a proper factor that cancel out any posible dimension in the equation. 

Through this process, some characteristic dimensionless numbers may appear, and they usually acquire some relevance depending on their physical significance. Let us consider the Navier Stokes equations for an incompressible fluid affected by a gravitational field $\vec{g}$:

\begin{equation}
\frac{\partial \vec{u}}{\partial t}+(\vec{u}\cdot \nabla) \vec{u}=-\frac{1}{\rho}\nabla P-\vec{g}+\nu\nabla^2\vec{u}.
\end{equation}

\noindent The characteristic dimensionless numbers of this general problem can be obtained easily by transforming (4) into its dimensionless form by considering the dimensionless variables $\vec{r'}=\vec{r}/L$, $t'=tU/L$, $\vec{u'}=\vec{u}/U$, $P'=P/\rho U^2$, so that we get:

\begin{equation}
\frac{\partial \vec{u'}}{\partial t'}+(\vec{u'}\cdot \nabla') \vec{u'}=-\nabla' P'-Fr^{-2} (\vec{g})+Re^{-1}\nabla^2\vec{u}.
\end{equation}

\noindent Where the desired non-dimensional parameters are the Reynolds and the Froude numbers, $Re=UL/\nu$ and $Fr=U/\sqrt{gL}$ respectively. But an alternative procedure to obtain dimensionless parameters without handling the equations of motion, which is more in the spirit of the Buckingham $\Pi$ theorem, is as follows.

\section{Buckingham $\Pi$ Theorem}

The general framework states that for a given physical problem with $(q_1,q_2,...,q_n)$ fundamental variables involved such that there exist a functional relationship of the form $f(q_1,q_2,...,q_n)=0$, there must exist $(n-j)$ dimensionless parameters called the $\Pi$ numbers, which can be constructed from combinations of the $n$ characteristic variables of the problem \cite{3}, where $j$ is the rank of the corresponding dimensional matrix. Tipically the number of fundamental dimensions or base units required to write the $n$ variables of the problem is the same as $j$. From the set of the $n$ variables one must choose $j$ variables as the "repeating variables" which must appear in all the nondimensional $\Pi$ numbers. It is important to remark that these repeating variables must have different dimensions between each other.

Now each $\Pi_i$ dimensionless number $(i=1,...,n-j)$ is formed from the product of the $j$ repeating variables and one of the $(n-j)$ remaining fundamental variables, where each one of the repeating variables must have an unknown exponent:

\begin{equation}
\Pi_i=V_{i} \prod_{k=1}^j V_{Rk}^{a_k}.
\label{eq:teopi2}
\end{equation}

Here $V_i$ is the fundamental variable asociated to $\Pi_i$ and it belongs to the set of the $(n-j)$ remaining variables. $V_{Rk}$ is the index over all the repeating variables. Finally $a_k$ represents the corresponding exponent to be determined in such a way that all the parameters $\Pi_i$ should be non dimensional. In this way is possible to rewrite  $f(q_1,q_2,...,q_n)=0$ as:

\begin{equation}
F(\Pi_1,\Pi_2,\ldots,\Pi_{n-j})=0.
\end{equation}

\section{Aplication to Rutherford scattering}

The best way to illustrate the power of the Buckingham $\Pi$ theorem is by an example. In this paper we choose a particular situation where dimensional analysis have not been used before, that is 
the deduction of the functional relation $s=s(\Phi)$ for Rutherford scattering.

So we first choose the set of fundamental variables of the problem as $(v,s,\phi,m_{\alpha}, F_e)$. Where $F_e$ the force between the heavy nucleus and the incident alpha particle for a typical distance. In this case $j=3$ and from the above set we choose $(v,s,F_e)$ to be the repeating variables. The theorem states that there are only two dimensionless parameters,  the first is the scattering angle itself since this is already a non-dimensional quantity $(\Pi_1=\Phi)$, and the second one can be founded from (6) using the three repeating variables choosen previously and $V_i=m_{\alpha}$:

\begin{equation}
\Pi_2=m_\alpha v^a s^b F_e^c,
\label{eq:param3}
\end{equation}

\noindent Which in terms of the three fundamental dimensions required in the problem turns out to be:

\begin{equation}
\Pi_2 = {M}^{1+c}L^{a+b+c}T^{-a-2c}
\end{equation}

The requirement of $\Pi_2$ to be dimensionless gives a simple linear system of equations for the unknown exponents $(a,b,c)$. The solution is quite straightforward and gives $a=2$, $b=-1$ and $c=-1$. Now replacing these values of the exponents into (8) we get:

\begin{equation}
\Pi_2=\frac{m_\alpha v^2}{s F_e}.
\label{eq:param6}
\end{equation}

\noindent Now notice that a tipical value for the electric force is given by:

\begin{equation}
F_e=\frac{kZZ' e^2}{s^2}.
\label{eq:param7}
\end{equation}

\noindent And replacing this into the expression for $\Pi_2$ gives:

\begin{equation}
\Pi_2=\frac{m_{\alpha} v^2 s}{k Z Z' e^2}
\end{equation}

\noindent Finally using (7) is easy to see that:

\begin{equation*}
\Pi_2=f(\Pi_1).
\end{equation*}

\noindent Then this can be used to obtain an expression for $s$, as follows:

\begin{equation}
s=\frac{k Z Z' e^2}{m_{\alpha} v^2}f(\Phi)
\end{equation} 

\noindent Which is in perfect accordance with (2).

\section{Conclusions}

Dimensional analysis and particulary the Buckingham $\Pi$ theorem is mainly used in fluid dynamics. However, in spite of this, in this paper we show a very simple and yet, powerfull use in a classical scattering problem, where it was shown that is posible to obtain a correct functional form of the impact parameter without using Newton Laws or any other conservation principle.


\section*{References}
\begin{thebibliography}{99}

\bibitem{1}  Weinberg S 1983 {\it Subatomic Particles} (New York: Scientific American Books Inc)
\bibitem{2}  Goldstein H, Poole C P and Safko J L 2002 {\it Classical Mechanics} (Addison Wesley)
\bibitem{3}  Kundu P and Cohen I 2002 {\it Fluid Mechanics} (Academic Press)
\bibitem{4} Cengel Y and Cimbala J 2009 {\it Fluid Mechanics with Student Resources} )(McGraw-Hill) 2nd edition.

\end{thebibliography}
\end{document}